\documentclass{aa}

\usepackage{graphicx}
\usepackage[varg]{txfonts}
\usepackage{amsmath}
\usepackage{verbatim}
\usepackage{latexsym}
\usepackage{multirow}

\begin{document}

\title{Image compression in local helioseismology}

\author{B. L\"optien\inst{1}
\and A.~C. Birch\inst{2}
\and L. Gizon\inst{1,2}
\and J. Schou\inst{2}}

\institute{Institut f\"ur Astrophysik, Georg-August Universit\"at G\"ottingen, 37077 G\"ottingen, Germany
\and Max-Planck-Institut f\"ur Sonnensystemforschung, Justus-von-Liebig-Weg 3, 37077 G\"ottingen, Germany}

\date{Received <date> /
Accepted <date>}

\abstract {Several upcoming helioseismology space missions are very limited in telemetry and will have to perform extensive data compression. This requires the development of new methods of data compression.}
{We give an overview of the influence of lossy data compression on local helioseismology. We investigate the effects of several lossy compression methods (quantization, JPEG compression, and smoothing and subsampling) on power spectra and time-distance measurements of supergranulation flows at disk center.}
{We applied different compression methods to tracked and remapped Dopplergrams obtained by the {\it Helioseismic and Magnetic Imager} onboard the {\it Solar Dynamics Observatory}. We determined the signal-to-noise ratio of the travel times computed from the compressed data as a function of the compression efficiency.}
{The basic helioseismic measurements that we consider are very robust to lossy data compression. Even if only the sign of the velocity is used, time-distance helioseismology is still possible. We achieve the best results by applying JPEG compression on spatially subsampled data. However, our conclusions are only valid for supergranulation flows at disk center and may not be valid for all helioseismology applications.}
{}


\maketitle


\section{Introduction}
Efficient compression of helioseismic data plays an important role in space missions, where it helps reduce the amount of telemetry required for transmitting data. The currently running {\it Helioseismic and Magnetic Imager}~\citep[HMI,][]{2012SoPh..275..229S} is not very limited in telemetry and relies on quantization and lossless compression only. On the other hand, the {\it Michelson Doppler Imager} (MDI) instrument~\citep{1995SoPh..162..129S} had to use extensive data compression. During the medium-$\ell$ program~\citep{1996kosovichev,1997SoPh..170...43K}, for example, the images were cropped, smoothed and subsampled, and quantized before applying lossless compression to the data.

Some upcoming missions, such as the {\it Solar Orbiter} mission~\citep{Yellowbook,Redbook,ISSI_Solar_Orbiter}, will also have a low bandwidth for transferring data. These missions can, of course, benefit from the experience in compression obtained by previous missions, in particular MDI. However, compression efficiency can certainly be improved, especially when using lossy compression algorithms. The behavior of lossy methods is complex and involves a trade-off between the compression efficiency and the amount of noise caused by the compression. The impact of compression artifacts might also change for different helioseismic measurements.

Like in many other cases, helioseismology can benefit from the experience with compression in Earth seismology. When analyzing seismic noise, it is common to use only the sign of the oscillations. This helps in the analysis because it removes seismic events with large amplitudes~\citep[e.g.,][]{1965Geop...30..665A,2013GeoJI.195.1811H}.

In this work, we give a first overview of lossy data compression in local helioseismology. We test the impact of compression on two of the most basic helioseismic measurements: the power spectrum and time-distance helioseismology of supergranulation at disk center. We start from Dopplergrams obtained by the HMI instrument, apply different compression methods (quantization, JPEG compression, and smoothing and subsampling) to the data and discuss their influence on the resulting power spectra and travel time maps. This allows us to give a first estimate on the quality and the efficiency of these compression methods.

\section{Compression schemes} \label{sect:comp}

\subsection{Quantization and Huffman encoding}
Quantization compresses data by reducing the number of bits per pixel used for storing the data. The number of bits per pixel determines how many different values $n$ the velocity can assume, e.g., for five bits per pixel $n=32$. Hence, quantization reduces the precision of the velocity. 

We apply quantization to the Dopplergrams by rounding the velocity to a number of possible values $n$ which are evenly distributed between the lowest and highest velocity in the time-series. The properties of the noise introduced by the quantization depends on $n$. For high $n$, this quantization noise is spatially uncorrelated and is roughly equivalent to white noise. Lower values of $n$ lead to a dependence of the quantization noise on the input velocities, causing sharp edges in the Dopplergrams. A special case is $n=2$; here it is only possible to distinguish between up- and downflows relative to the mean velocity.

Additional compression can be achieved by combining quantization with a lossless compression algorithm. Quantization only decreases the number of bits per pixel that are saved to a fixed value. However, using a fixed number of bits per pixel is not very efficient, since the histogram of the velocities does not correspond to a uniform distribution. Velocities with low amplitudes are much more frequent than those with high amplitudes. We apply Huffman encoding~\citep{Huffman}, an entropy-based lossless compression algorithm which uses a non-constant number of bits per pixel for storing the data. More frequent values of the velocity are saved using fewer bits per pixel than less frequent ones.

The efficiency of Huffman compression depends on the statistics of the input data. The more information there is in the data, the larger the file size of the compressed data. Since the velocities in the Dopplergrams are spatially and temporally correlated, we decrease the amount of information significantly by predicting the velocity of each pixel from the surrounding pixels both in time and space using linear regression. Standard Huffman encoding does not allow us to reach file sizes smaller than one bit per pixel because every pixel is saved separately. We avoid this by concatenating the velocities from three consecutive pixels into one symbol before using the Huffman compression.

We test quantization for different values of $n$, ranging from 256 to two. In reality, the velocity would probably not be rounded to a fixed number of possible values between the lowest and highest velocity in the image, but it would be truncated to a fixed precision. This is better if e.g., cosmic rays or gradients in velocity across the image are present.

\subsection{JPEG compression}
Another option for compressing data is to truncate the coefficients of some spatial transformation of the data. This leads to a loss of information about small spatial scales. A common example for such a method is JPEG (Joint Photographic Experts Group) compression~\citep{Wallace1992}.

JPEG compression divides the data into blocks with a size of $8\times 8$ pixels, then a discrete cosine transform (DCT) is applied to each block. The coefficients of the DCT are truncated depending on a quality factor $q$ that has to be selected as an input parameter (between 0 and 100, with a lower factor meaning a higher compression ratio) and compressed using Huffman encoding. JPEG compression does not benefit from combining it with quantization. The sharp edges in the Dopplergrams introduced by the quantization affect the higher coefficients of the DCT and increase the file size.

We use the standard JPEG compression algorithm implemented in IDL\footnote{IDL (Interactive Data Language) is a product of EXELIS Visual Information Solutions, http://www.exelisvis.com/} and test different values of the quality factor $q$ (between 5 and 100).

\subsection{Smoothing and subsampling}
Another method for reducing the size of the data is decreasing the spatial resolution by subsampling the data. Subsampling decreases the Nyquist wavenumber of the data. In our case, we use only $2\times 2$ subsampling on HMI data near disk center, meaning that the Nyquist wavenumber is still at $kR_\odot \approx 3100$. So, aliasing only has a small influence on the power spectrum and it is sufficient to convolve the Dopplergrams with a narrow Gaussian ($\sigma =0.4$~pixels) to remove artifacts resulting from aliasing. After subsampling, we apply one of the two compression methods described above to the Dopplergrams.

\section{Results}

\subsection{Input observations}
\begin{figure}
\centering
\includegraphics[width=\columnwidth]{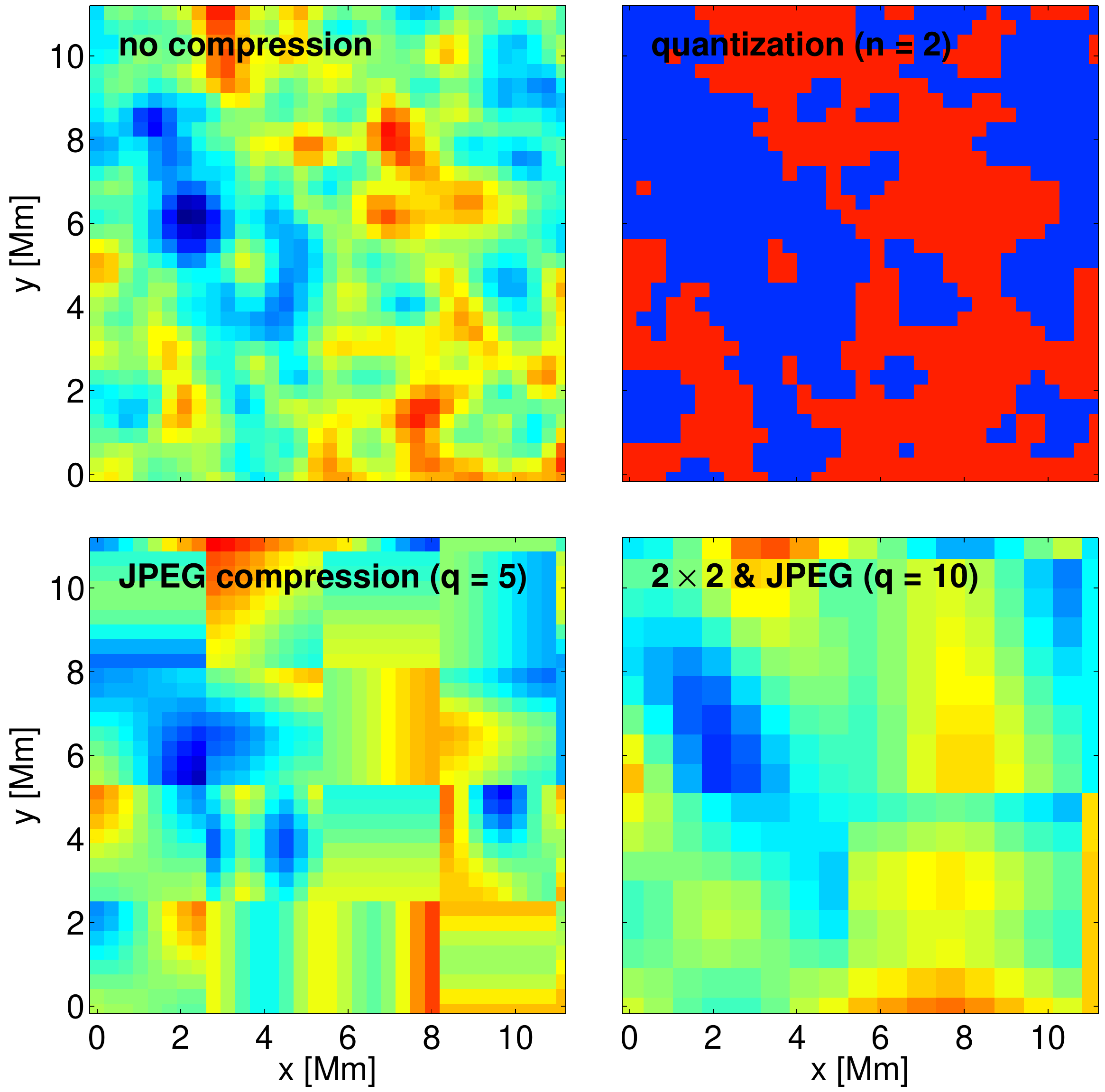}
\caption{Examples of uncompressed and compressed Dopplergrams. {\it Top left}: uncompressed data, {\it top right}: quantization (two velocity bins), {\it bottom left}: JPEG compression (quality = 5), {\it bottom right}: $2\times 2$ subsampling combined with JPEG compression with a quality of 10. The artifacts resulting from the compression are clearly visible in the Dopplergrams. Quantization causes sharp edges in the image; if $n=2$, only upflows and downflows can be separated. JPEG compression leads to blocks in the image with a size of $8\times 8$~pixels, corresponding to $2.78$~Mm or $5.56$~Mm, if $2\times 2$ subsampling is used. The images shown represent only a small part of the full size of the Dopplergrams ($178 \times 178$~Mm).}
\label{fig:dopplergrams}
\end{figure}
Our analysis is based on Dopplergrams for the quiet Sun provided by the HMI instrument. This instrument obtains filtergrams at six wavelength positions around the Fe I 6173 \AA \ line and transmits them to Earth, where the observables, e.g., Dopplergrams, are computed. Since HMI is basically unlimited in telemetry, the raw images that are downlinked by HMI have a size of $\sim 7.1$ bits per pixel. Hence, the precision of the resulting Dopplergrams is far beyond the photon noise level.

We use twenty time-series of Dopplergrams obtained in May 2010 that are tracked and remappped at the equator corresponding to the Mt Wilson 1982/84 differential rotation rate~\citep{1984SoPh...94...13S} around the central meridian. Each of the time-series has a length of eight hours and a size of $178\times 178$ Mm ($512\times 512$ pixels, spatial resolution: $0.348$~Mm). The tracked and remapped Dopplergrams are saved in single precision (32 bits per pixel). In the next step, we apply the various compression schemes described in Sect.~\ref{sect:comp} to the data. We study a broad range of parameters (velocity bins $n$ and quality factor $q$), but the results we present in the following sections are mostly for the highest compression factors since these show the strongest influence of compression.

Examples of extremely compressed Dopplergrams are shown in Figure~\ref{fig:dopplergrams}. They clearly show the artifacts resulting from the compression. Quantization reduces the number of possible values of the velocity, causing sharp edges in the Dopplergrams, and JPEG compression leads to artifacts in the image based on the blocks used for the JPEG compression.

\subsection{Influence on helioseismic power spectra}
\begin{figure}
\centering
\includegraphics[width=\columnwidth]{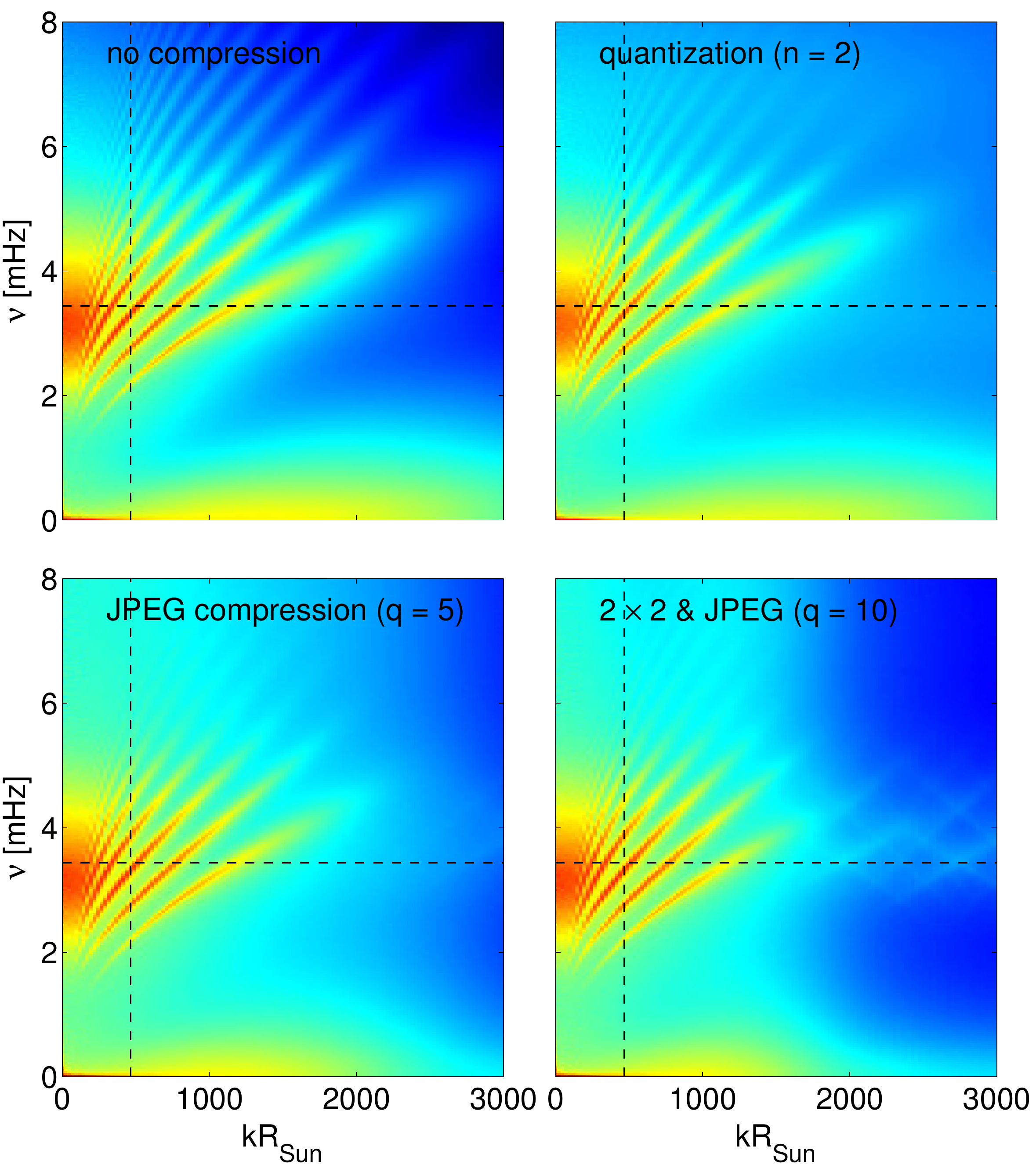}
\caption{Compression artifacts in azimuthally averaged power spectra. {\it Top left}: Uncompressed data, {\it top right}: quantization (two velocity bins), {\it bottom left}: JPEG compression (quality = 5), {\it bottom right}: $2\times 2$ subsampling combined with JPEG compression with a quality of 10. All four power spectra are plotted using the same logarithmic color-scale (red corresponds to high power, blue to low power) and are averages from twenty time-series, each of them having a length of eight hours. We normalized the bit-truncated Dopplergrams to have the same variance as the uncompressed data before computing power spectra. All compression methods slightly decrease the power of the modes and add additional noise to the power. The noise caused by quantization is almost flat, the noise caused by JPEG compression exhibits a more complex behavior, including several ridges appearing at high wavenumbers and a reduction of the power arising from granulation. In Figure~\ref{fig:power_spectra_cut}, 
we 
show cuts through the power spectra along the dashed lines.}
\label{fig:power_spectra}
\end{figure}
\begin{figure}
\resizebox{\hsize}{!}{\includegraphics{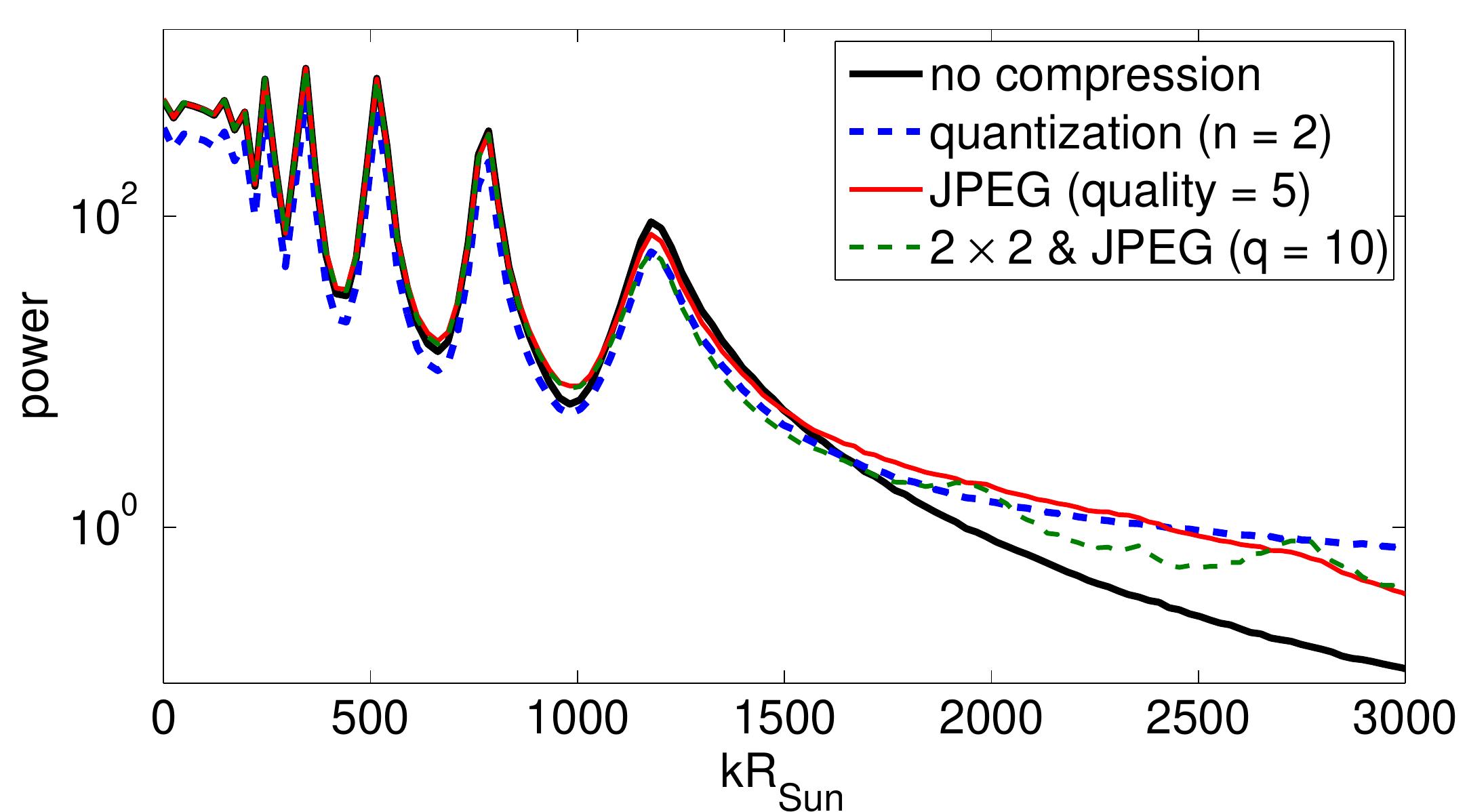}}
\resizebox{\hsize}{!}{\includegraphics{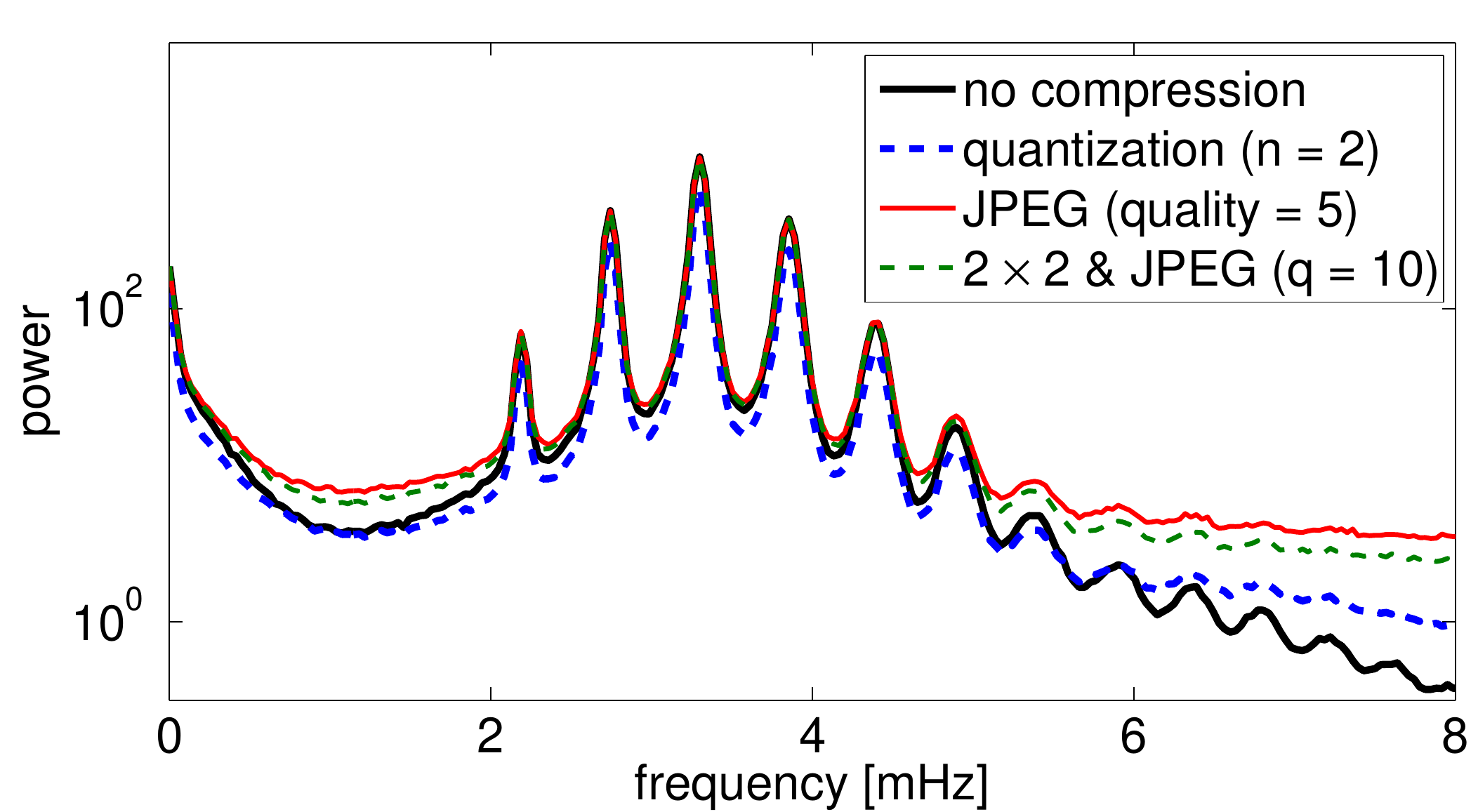}}
\caption{Cuts through the azimuthally averaged power spectra shown in Figure~\ref{fig:power_spectra} (cut along the dashed lines). {\it Top}: power at a constant frequency, $\nu=3.4$ mHz, {\it bottom}: power at a constant wavenumber, $kR_\odot=466$. Four configurations are shown. {\it Thick solid black curve}: no compression is applied to the data, {\it thick dashed blue curve}: the Dopplergrams are compressed using quantization (two velocity bins), {\it solid red curve}: JPEG compression (quality = 5), {\it dashed green curve}: $2\times 2$ subsampling and JPEG compression with a quality of 10. The power shown is an average of twenty time-series, each of them having a length of eight hours. We normalized the bit-truncated Dopplergrams to have the same variance as the uncompressed data before computing power spectra. All compression methods slightly decrease the power of the modes and add additional noise to the power.}
\label{fig:power_spectra_cut}
\end{figure}

The most basic tool in helioseismology is the power spectrum as a function of wavenumber and frequency. In Figs.~\ref{fig:power_spectra} and~\ref{fig:power_spectra_cut}, we compare azimuthally averaged power spectra for uncompressed and examples of extremely compressed data (quantization with two bins in velocity, JPEG compression with a quality factor of five, and JPEG compression with $q=10$ applied to $2\times 2$ subsampled data).

Even for high compression factors, the p- and f-modes are still clearly visible in the power spectrum. However, depending on the compression factor, the compression decreases the power of the oscillations and adds noise. The characteristics of the compression noise depend on the method. Quantization adds an almost flat background noise to the power spectrum. The noise caused by the JPEG compression exhibits a more complicated shape. For very low quality factors, several ridges appear at high wavenumbers, depending on the spatial resolution of the data. In addition, the power arising from granulation is reduced significantly.

\subsection{Influence on supergranulation travel times}
\begin{figure*}
\centering
\includegraphics[width=17cm]{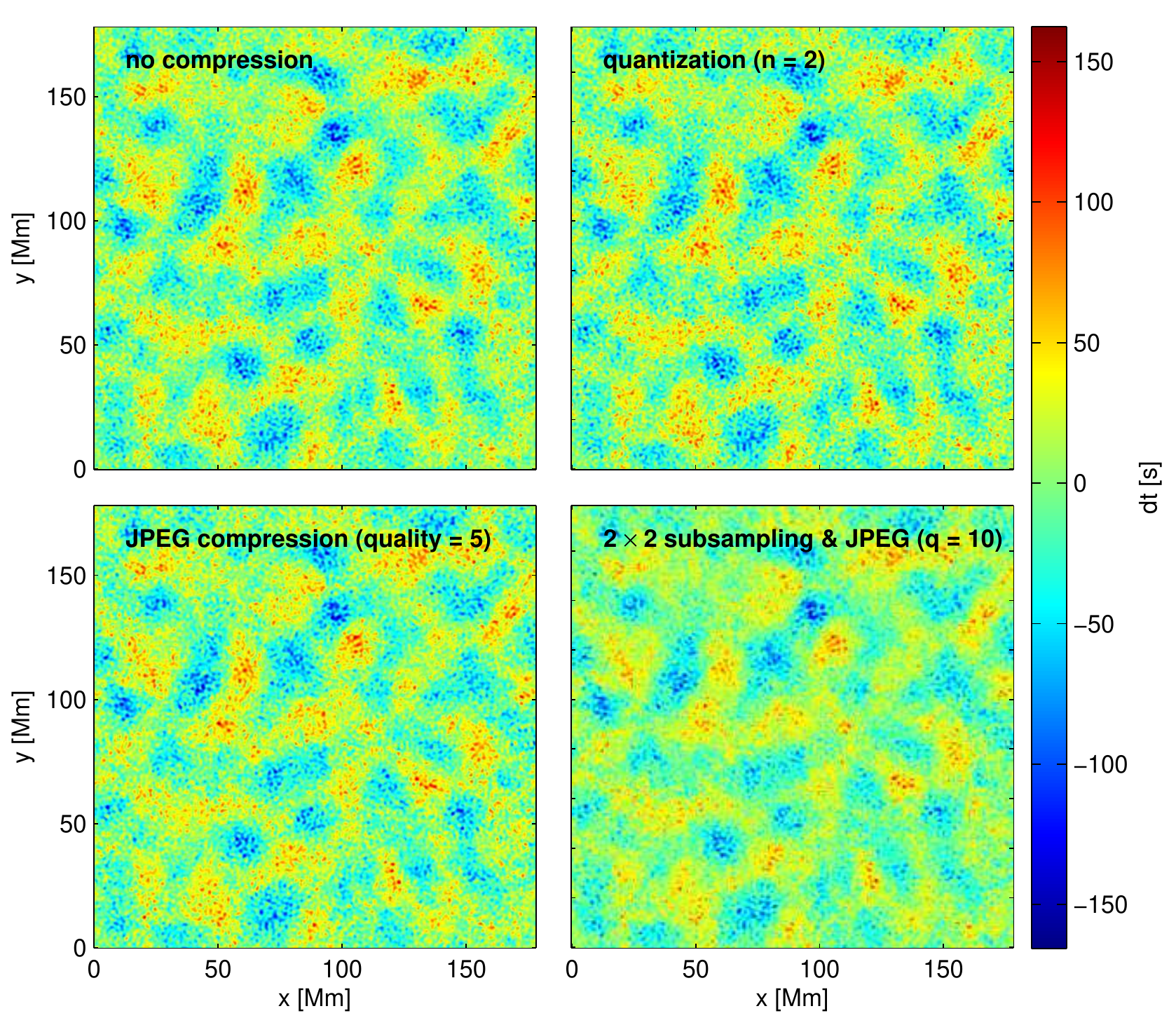}
\caption{Examples of travel time maps of supergranulation (f-mode, center-to-annulus geometry, annulus size $13.4$~Mm, outward minus inward travel times) computed from compressed and uncompressed Dopplergrams. {\it Top left}: Uncompressed data, {\it top right}: quantization (two velocity bins), {\it bottom left}: JPEG compression (quality = 5), {\it bottom right}: $2\times 2$ subsampling combined with JPEG compression with a quality of 10. The travel times clearly show the supergranulation pattern. They correspond to the divergence of horizontal flows, with negative values indicating outflows and positive values showing inflows.}
\label{fig:tt}
\end{figure*}
We evaluate the influence of the compression methods described in section~\ref{sect:comp} on time-distance helioseismology~\citep{1993Natur.362..430D} by computing travel time maps for the f-mode using point-to-annulus geometry (annulus size $13.2$~Mm) and by taking the difference in travel times between outward-going and inward-going waves. These travel times are sensitive to the horizontal divergence of near-surface flows~\citep{2000SoPh..192..177D}, such as supergranulation. Examples of the resulting travel time maps are shown in Figure~\ref{fig:tt}.

Although the compression introduces additional noise, the travel time maps for quantization and JPEG compression are almost indistinguishable from the uncompressed case. The correlation between uncompressed and compressed travel times is only for extreme compression ratios smaller than $0.99$ (see Table~\ref{tab:tt}). There is also no apparent correlation of the noise caused by compression with the travel times.
\begin{table*}
\caption{Properties of travel time maps (center-to-annulus geometry, outward minus inward travel times) computed from compressed data. We evaluate the quality of the compression for selected compression efficiencies (number of velocity bins $n$ or quality factor $q$) by computing the correlation of the travel times with the travel times computed from uncompressed data, the signal-to-noise ratio (S/N) of the travel times for two wavenumbers ($kR_\odot=122$ and $kR_\odot =221$) and the file size of the Dopplergrams in bits/pixel (relative to the data with full spatial resolution) after applying the compression. The $2\times 2$ subsampled data were interpolated back to the original resolution using Fourier interpolation for computing the correlations. Before determining the S/N, we averaged the power of the travel times from $kR_\odot=98 - 147$ and $kR_\odot=196-245$, respectively. Our measurement of the S/N for the uncompressed case has an uncertainty of about 2. The other error 
estimates are for differential S/N measurements relative to the uncompressed case.}
\label{tab:tt}
\centering
\begin{tabular}{p{3cm}lllll}
\hline\hline
compression method & parameter & Correlation & S/N at $kR_\odot=122$ & S/N at $kR_\odot=221$ & file size [bits/pixel]\\
\hline
no compression & & 1 & $40\pm 2$ & $8.0\pm 0.3$ & $32$\\
quantization & $n=2$ & $0.965$ & $35.1\pm 0.8$ & $6.5\pm 0.1$ & $0.71$\\
 & $n=4$ & $0.975$ & $36.8\pm 0.7$ & $6.81\pm 0.09$ & $0.72$\\
 & $n=8$ & $0.998$ & $39.8\pm 0.2$ & $7.84\pm 0.03$ & $0.92$\\
JPEG compression & $q = 5$ & $0.957$ & $31\pm 1$ & $5.8\pm 0.2$ & $0.23$\\
 & $q = 10$ & $0.991$ & $38.2\pm 0.4$ & $7.50\pm 0.08$ & $0.43$\\
 & $q = 20$ & $0.998$ & $39.2\pm 0.3$ & $7.91\pm 0.03$ & $0.71$\\

$2\times 2$ subsampling and & $n=2$ & $0.948$ & $32\pm 1$ & $5.9\pm 0.2$ & $0.23$ \\
quantization & $n=4$ & $0.960$ & $34\pm 1$ & $6.2\pm 0.2$ & $0.24$\\
 & $n=8$ & $0.989$& $37.9\pm 0.6$ & $7.36\pm 0.09$ & $0.30$\\
 
$2\times 2$ subsampling and & $q=5$ & $0.612$& $7\pm 2$ & $0.8\pm 0.3$& $0.06$\\
JPEG compression & $q=10$ & $0.927$ & $27\pm 2$ & $4.6\pm 0.2$ & $0.12$\\
 & $q=15$ & $0.972$& $34.2\pm 0.9$ & $6.0\pm 0.2$ & $0.17$\\
\hline
\end{tabular}
\end{table*}

When deciding on a compression method, two things are important: the quality of the compressed data and the compression efficiency. These depend on the parameters of the compression (the number of velocity bins $n$ of the quantization or the quality factor $q$ of the JPEG compression). Determining the compression efficiency is straightforward; here we determine the quality of the compressed data by comparing the signal-to-noise ratio (S/N) with the uncompressed data.

The S/N of travel time maps is defined as the difference between the power of the travel times computed from the observations and a noise model divided by the noise~\citep{2004ApJ...614..472G,Fournier2014}.
\begin{figure*}
\centering
\includegraphics[width=17cm]{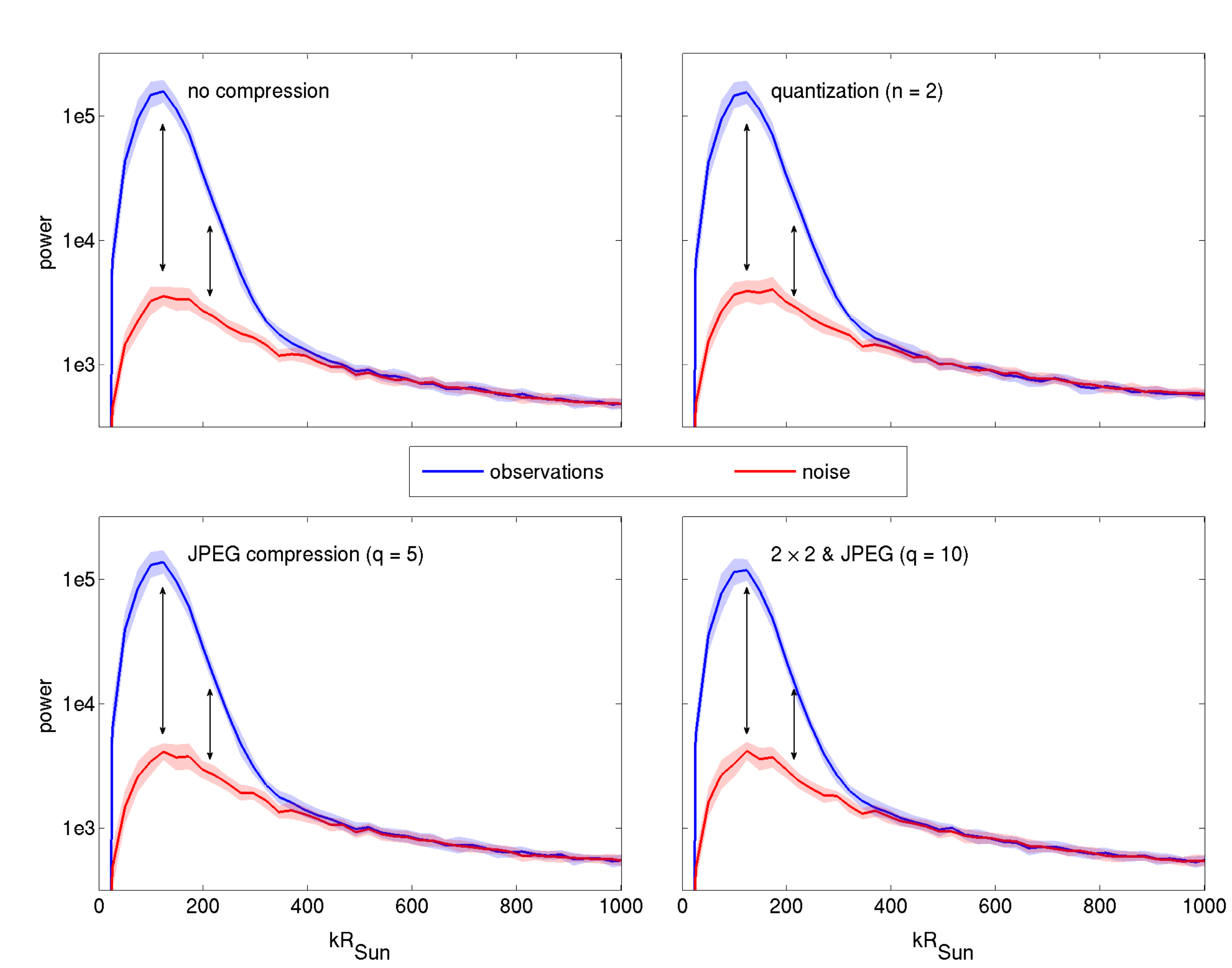}
\caption{Azimuthally averaged power of travel times (center-to-annulus geometry, outward minus inward travel times) derived from observations ({\it blue curve}) and a noise model ({\it red curve}). {\it Top left}: Uncompressed data, {\it top right}: quantization (two velocity bins), {\it bottom left}: JPEG compression (quality = 5), {\it bottom right}: $2\times 2$ subsampling combined with JPEG compression with a quality of 10. The power shown is an average from twenty time-series, each of them having a length of eight hours. The shaded areas show the $1-\sigma$ scatter of the individual realizations. The compression slightly decreases the signal and increases the noise. The vertical arrows denote the wavenumbers used for determining the S/N in Figure~\ref{fig:tt_qual} and Table~\ref{tab:tt}.}
\label{fig:tt_power}
\end{figure*}
Both the observations and the noise model are affected by compression. As can be seen in Figure~\ref{fig:tt_power}, the compression slightly decreases the power caused by the supergranulation and enhances the effect of the realization noise. This decreases the S/N.

\begin{figure}
\centering
\includegraphics[width=\columnwidth]{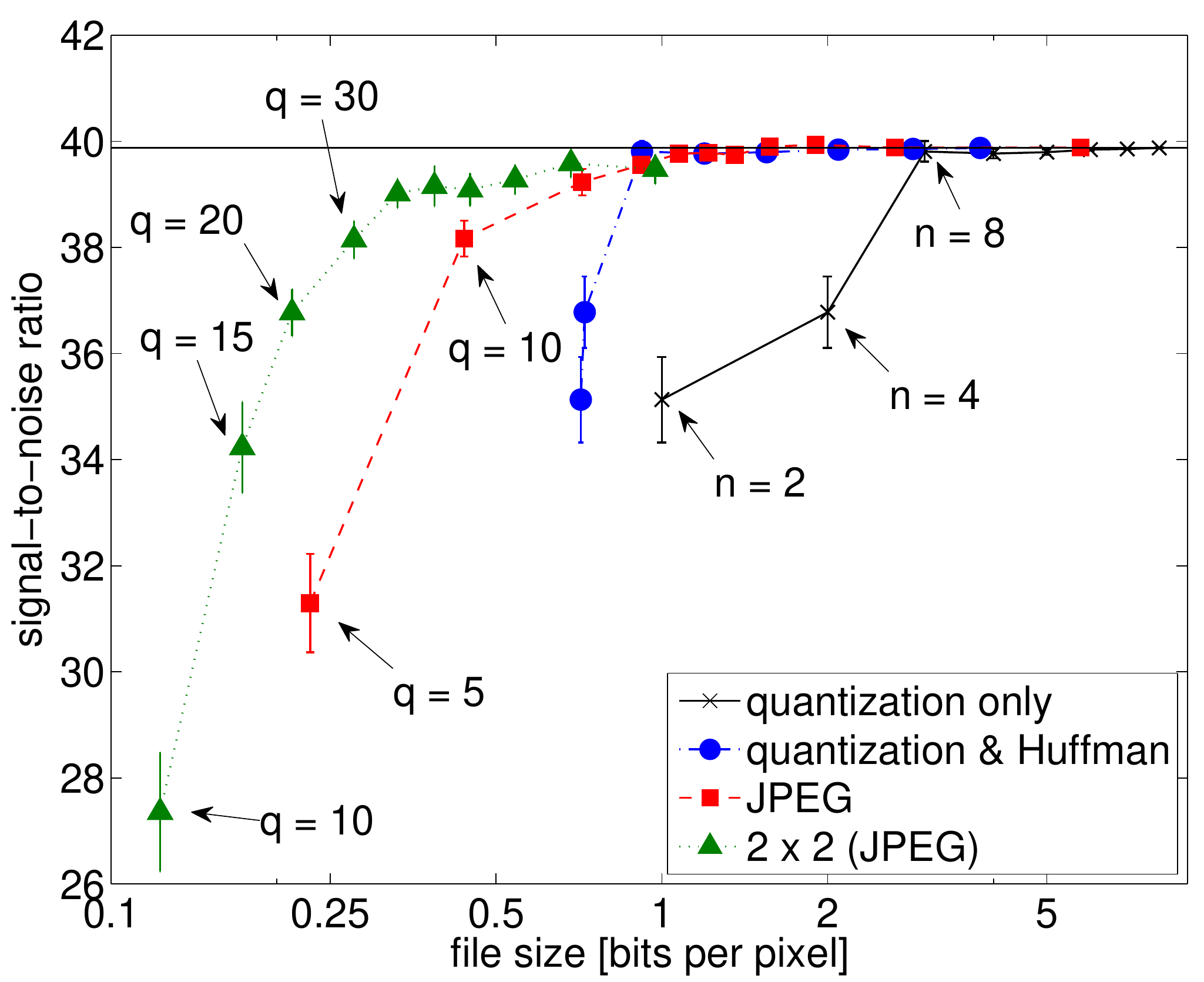}
\caption{S/N of travel times (center-to-annulus geometry, outward minus inward travel times) at a wavenumber of $kR_\odot=98 - 147$ as a function of the file size in bits per pixel (relative to the data with full spatial resolution). We show data for quantization and JPEG compression using both the full spatial resolution and $2\times 2$ subsampled data. {\it Black crosses}: quantization only (full resolution), {\it blue circles}: quantization and Huffman compression (full resolution), {\it red squares}: JPEG compression (full resolution), {\it green triangles}: JPEG compression and $2\times 2$ subsampling. We vary the number of possible values for the velocity $n$ for the quantization and the quality factor $q$ of the JPEG compression (as indicated by the arrows). We do not show the S/N for JPEG compression with $q=5$ applied to $2\times 2$ subsampled data because it is extremely low (S/N = 7). The S/N shown here is an average computed from twenty realizations. The horizontal line 
shows the S/N of the uncompressed data. Before computing the S/N, we averaged the power of the travel times from $kR_\odot=98 - 147$. Our measurement of the S/N for the uncompressed case has an uncertainty of about 2. The error bars show error estimates for differential S/N measurements relative to the uncompressed case.}
\label{fig:tt_qual}
\end{figure}

We determine the S/N at two wavenumbers which correspond to the length scale of supergranulation ($kR_\odot=122$ and $kR_\odot=221$, indicated by the vertical arrows in Figure~\ref{fig:tt_power}). In Table~\ref{tab:tt} and in Figure~\ref{fig:tt_qual}, we show the resulting S/N and the file sizes for the different compression methods while varying the parameters of the compression (the number of velocity bins $n$ for the quantization and the quality factor $q$ of the JPEG compression). Since the behavior of the compression methods is very similar at these two wavenumbers, we show only the S/N at $kR_\odot=122$ in the Figure.

All compression methods decrease the file size significantly; a compression to a file size of one bit per pixel is possible without any major influence on the S/N. However, there are large differences between the individual methods. In general, JPEG achieves a better S/N than the combination quantization and Huffman encoding, especially for high compression factors. A quality factor of ten leads to a S/N of $38$ at $kR_\odot = 122$ which is comparable to the S/N of the uncompressed data (S/N = $40$). This corresponds to a file size of $0.43$~bits per pixel. Additional compression can be achieved by using JPEG compression on $2\times 2$~subsampled data. This is the best method tested in this study. No matter which S/N is required (except for a S/N > $39.5$), this compression scheme leads to the smallest file size. A further reduction of the spatial resolution is not advisable, using $4\times 4$ subsampling significantly reduces the S/N (the maximum S/N is 20). The spatial resolution of $4\times 4$ 
subsampled data is $4\times 0.348 = 1.392$~Mm. This is comparable to the typical wavelength of the f-mode ($\sim 4-8$~Mm); $4\times 4$ subsampling might work better for p-modes which have at the same frequency a larger wavelength than the f-mode.

\section{Discussion and conclusion}

Probing supergranulation flows at disk center seems to be very robust to data compression. However, the good performance of the compression is probably not valid for helioseismology in general. There are helioseismic measurements with a much lower S/N than the simple measurements presented in this paper, for example, weak and deep flows and helioseismology at high latitudes. Most likely, these are much more sensitive to data compression. In addition, center-to-limb effects could have a strong effect here. Subsampling and JPEG compression should presumably both be affected by foreshortening when applied close to the limb. The effect of compression on other measurement techniques also needs to be studied. For example, the reduction of the power arising from granulation by the JPEG compression indicates that granulation tracking will be affected by this compression method. Hence, more work is needed in order to derive a strategy for data compression that can be applied to all sorts of 
helioseismic measurements.

While the excellent performance of the compression for travel time measurements might be surprising at first, it can be explained by looking at the power spectra. Most of the compression artifacts are located at higher wavenumbers than used in time-distance helioseismology. This means that the filtering applied for the travel time measurements removes most of the compression noise. The noise at the location of the modes is small compared to the power of the oscillations. Hence, it has only minor influence on travel time measurements.

Based on the results shown in Figure~\ref{fig:tt_qual}, subsampling in combination with JPEG compression seems to be the best of the methods tested here for compressing data for local helioseismology. These results are, of course, limited to HMI. Helioseismic analyses require a minimum spatial resolution, so there is an upper limit on the amount by which the data can be subsampled, depending on the spatial resolution of the instrument and the wavelength of the target waves. For HMI, $2\times 2$ subsampling seems to be a good trade-off between file size and image quality. In combination with JPEG compression (with varying quality factor), this reduces the file size significantly. Which compression factor can be achieved depends on the science goal of the analysis. If a low S/N is sufficient (e.g., for a statistical analysis of supergranulation), the data can be compressed down to $\sim 0.15$~bits per pixel (relative to the full spatial resolution). Even if a high S/N is required, a compression to 
$\sim 0.3$~bits per pixel is possible without decreasing the quality of the data too much.

The file sizes can probably be decreased even further. The compression methods presented in this study are very simple and probably not optimal. JPEG compression, for example, is designed to work for photos, not for scientific applications. The efficiency of JPEG could probably also be increased by including the time-domain in the transformation instead of only using the spatial dimensions. This also increased the efficiency of the Huffman compression.

\begin{acknowledgements}
We are grateful to Jan Langfellner for providing the tracked and remapped Dopplergrams and the noise cubes. We acknowledge support from Deutsche Forschungsgemeinschaft (DFG) through SFB 963/1 "Astrophysical Flow Instabilities and Turbulence" (Project A1). Support was also provided by European Union FP7 projects SPACEINN and SOLARNET.
 
\end{acknowledgements}

\bibliographystyle{aa} 
\bibliography{literature} 

\end{document}